\documentclass[aps,reprint,amsmath,amssymb]{revtex4-2}
\usepackage{graphicx,subfigure}
\usepackage{epstopdf}
\usepackage{upgreek}
\usepackage{bm}
\usepackage{dcolumn}
\usepackage{epstopdf}
\usepackage{color}

\makeatletter
\def\btt#1{\texttt{\@backslashchar#1}}%
\DeclareRobustCommand\bblash{\btt{\@backslashchar}}%
\makeatother

\makeatletter
\def\btt#1{\texttt{\@backslashchar#1}}%
\DeclareRobustCommand\bblash{\btt{\@backslashchar}}%
\makeatother
\begin{document}
\title{Mixing twist-bend and ferroelectric nematic liquid crystals}
\date{\today}
\author{Abinash Barthakur$^{1}$, Jakub Karcz$^{2}$, Przemyslaw
 Kula$^{2}$ and Surajit Dhara$^{1}$}
\email{surajit@uohyd.ac.in} 
\affiliation{$^{1}$School of Physics, University of Hyderabad, Hyderabad-500046, India\\
$^{2}$Institute of Chemistry, Faculty of Advanced Technologies and Chemistry, Military University of Technology, Warsaw, Poland  
}

\begin{abstract}
 Twist-bend (N\textsubscript{tb}) and ferroelectric  (N\textsubscript{F}) nematic liquid crystals exhibit several novel effects and new physical properties. The question of what happens in binary mixtures is interesting as a matter of curiosity and pure science. 
  Here, we report experimental studies on the phase diagram and some physical properties of binary mixtures of the above-mentioned nematic liquid crystals.  Both N-N\textsubscript{tb} and N-N\textsubscript{F} phase transition temperatures and the corresponding enthalpies decrease significantly and eventually, these transitions disappear in some intermediate compositions. Temperature-dependent birefringence above the N-N\textsubscript{tb} phase transition temperature shows critical behaviour, and the critical range of the tilt fluctuations becomes wider in the mixtures. The magnitudes and the temperature dependence of the splay elastic constant of the mixtures' high-temperature nematic (N) phase strikingly differ from that of the pristine twist-bend and ferroelectric nematic liquid crystals. 
\end{abstract}
  
  \preprint{HEP/123-qed}
\maketitle
\section{Introduction}
The discovery of new nematic liquid crystals with nanoscale modulation of director (twist-bend nematic, N\textsubscript{tb})~\cite{VPN,MCE1,MCE2,Dch, MCO, VBO,LBE,GPA} and spontaneous electric polarisation (ferroelectric nematic, N\textsubscript{F})~\cite{RJM1,RJM2,NHI,MAE,AMA,NSE,XCH} has created immense interest. They have lower symmetry than conventional apolar nematic  (N) liquid crystals. The twist-bend phase was discovered in bend-core compounds possessing two rigid rod-like mesogenic units connected by a flexible chain~\cite{MCE2}. The ferroelectric nematic was discovered in compounds composed of molecules with very high axial dipole moment~\cite{RJM1,NHI}. Interestingly, both these phases were predicted theoretically much before their experimental discovery~\cite{MBO,IDO}. Particularly, the ferroelectric nematic was envisaged more than a century ago by Max Born~\cite{MBO}. These nematic liquid crystals exhibit several new physical properties which are of great fundamental and practical interest. For example, N\textsubscript{tb} shows a pseudolayer structure~\cite{VBO} and the bend elastic constant $K_{33}$ in the high-temperature nematic phase is very small~\cite{VBO}. On the other hand $K_{11}$  is unusually small near the N-N\textsubscript{F} transition~\cite{MAE}. The nematic twist-bend phase provides a compressional elastic constant and smectic-like rheology~\cite{Pra1,Pra2}.  N\textsubscript{F} phase shows striking electrooptics and a large spontaneous polarisation along the director and giant electrohydromechanical effects~\cite{Bart,MTM1,RB,MTM2,Pra3}. Ferroelectric nematic LCs are promising for applications due to their low threshold field and very fast switching response time.\\

Distinct molecular structures and new physical properties of these nematic liquid crystals naturally generate curiosity about the mixtures and their physical properties. Experimental investigation on mixtures of compounds often leads to new phase transitions and physical properties~\cite{PRAT,BKK}. Recently some interesting results have been reported in the binary mixture of paraelectric and ferroelectric phases of 4-((4-Nitrophenoxy)carbonyl)phenyl 4-methoxy-2-methoxybenzoate (RM734)  and 2,3$^{'}$,4$^{'}$,5$^{'}$- tetrafluoro-[1,1$^{'}$-biphenyl]-4-yl 2,6-difluoro-4-(5-propyl-1,3-dioxan- 2yl)benzoate (DIO) nematic liquid crystals ~\cite{CLARK}. Naturally, a question arises: Are two new nematics ($N_{tb}$ and $N_{F}$) having distinct molecular structures and local symmetries miscible? How do physical properties change with the composition? To address these questions 
we prepared a binary phase diagram by mixing a twist-bend nematic and ferroelectric nematic liquid crystal. We systematically investigated phase transitions and measured some properties like enthalpies, birefringence, dielectric and splay elastic constant as a function of temperature. 
\begin{figure}[b]
\centering
\includegraphics[scale=0.15]{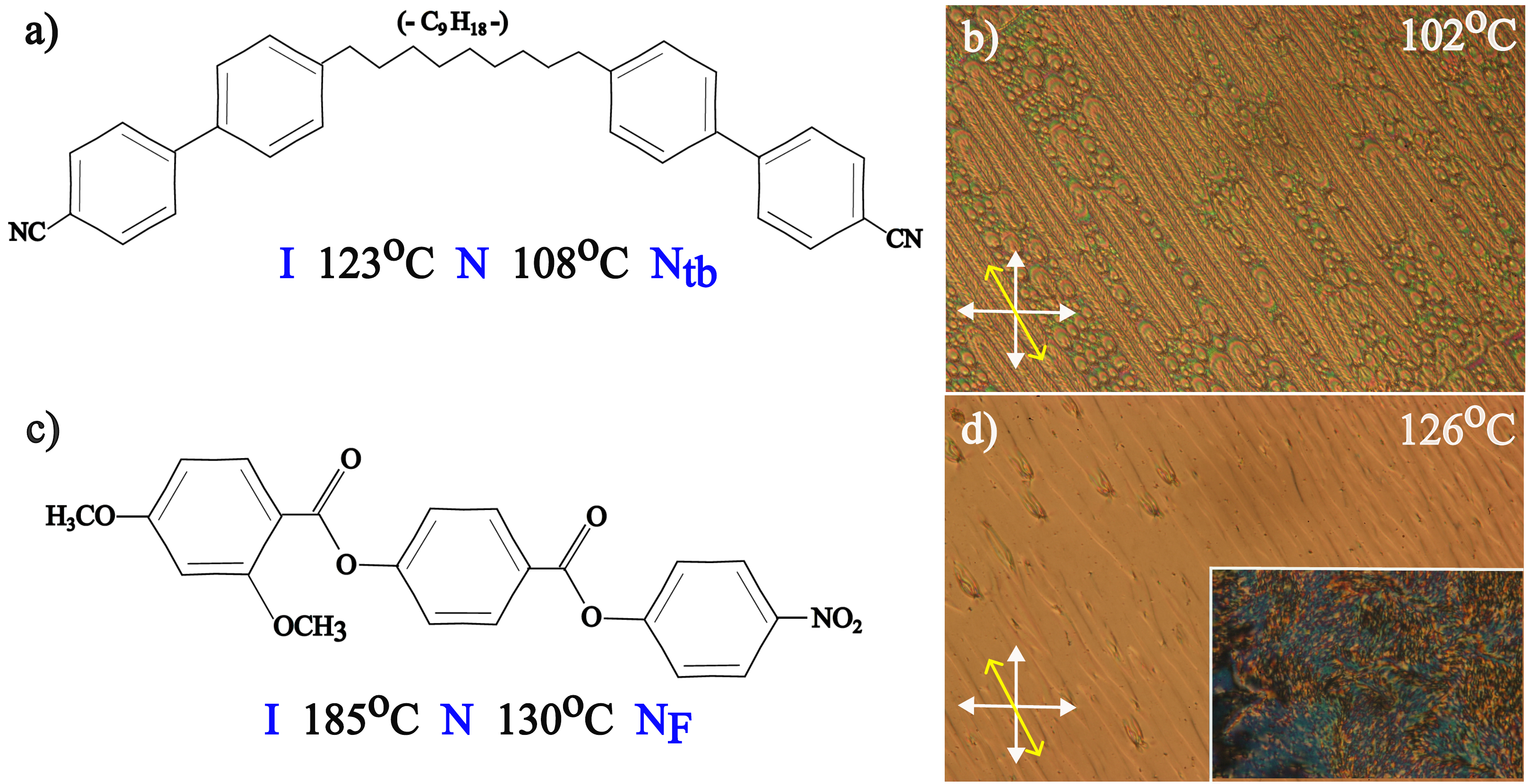}
\caption{ (a) Molecular structures and phase transition temperatures of (a) CB9CB and (c) RM734 compounds. Textures of  (b) N\textsubscript{tb}  and (d) N\textsubscript{F} phases in planar cells (10$\mu$m). The width of the image is 0.7 mm. Inset shows texture in an untreated cell.}
\label{fig:figure1} 
\end{figure}

\begin{figure*}
\centering
\includegraphics[scale=0.7]{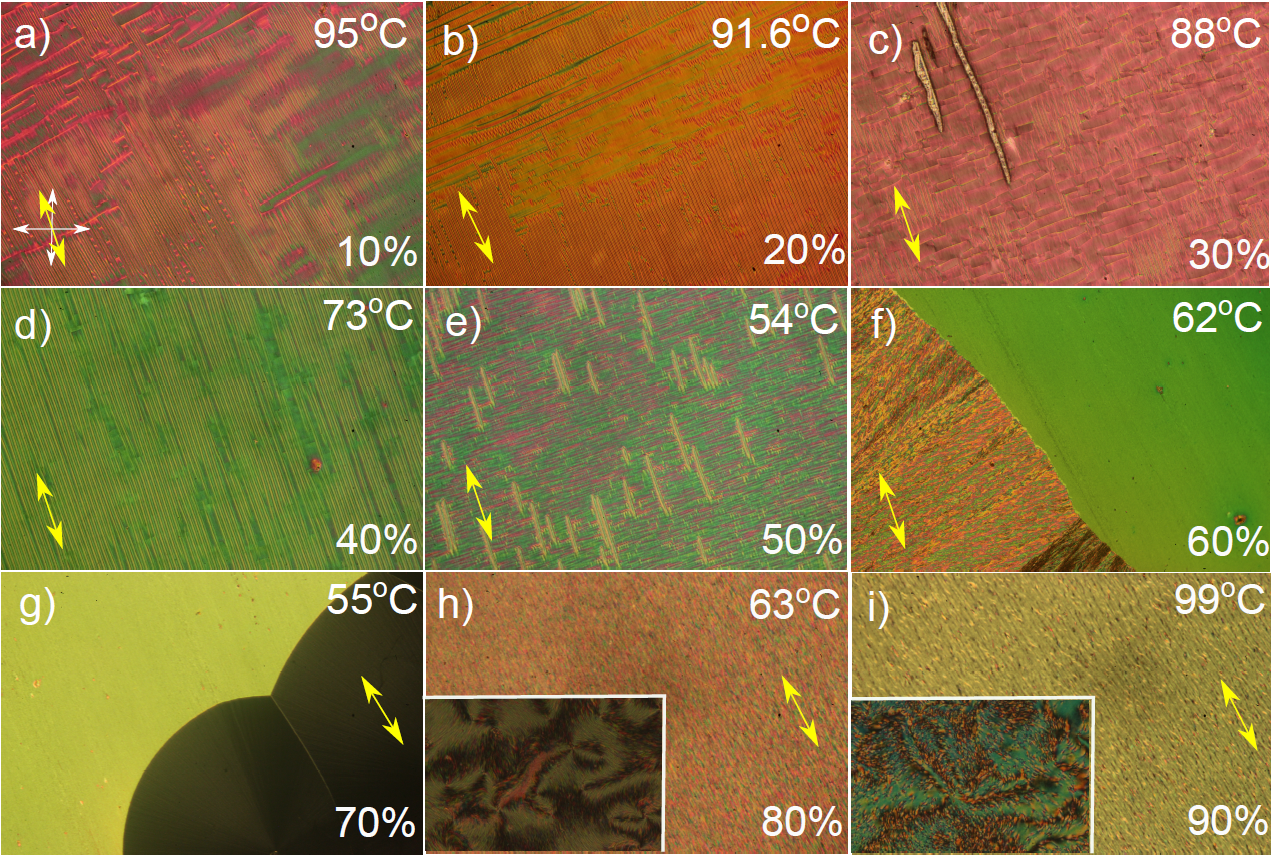}
\caption{Representative polarising optical microscope (POM) textures of a few mixtures at different temperatures.  Double-headed yellow arrows indicate the rubbing direction. Wt\% of RM734 and the temperatures are marked on each image. (a,b,c,d,e) N\textsubscript{tb} phase,  (f,g) N to crystal transition,  (h,i) N\textsubscript{F} phase.  Width of an image is 1.3 mm. Insets show textures in untreated cells.}
\label{fig:figure2} 
\end{figure*}
  
\section{Experimental}
Both  liquid crystals, 1\textsuperscript{''},7\textsuperscript{''}-bis(4-cyanobiphenyl-4\textsuperscript{'}-yl)nonane (CB9CB) and  RM734  were synthesised in our laboratory~\cite{Pra2,Pra3}. Molecular structures and the phase transition temperatures are shown in Fig.\ref{fig:figure1}(a) and Fig.\ref{fig:figure1}(c), respectively.  CB9CB molecules are bent-shaped due to the flexible odd-numbered hydrocarbon link  and RM734 molecules are wedged-shaped because of the lateral methoxy group~\cite{Alenka2021}. 
At room temperature both the compounds are in powder form. Appropriate weights of each material were put together in a conical glass cup and heated to 190\textsuperscript{o}C.  The sample was mixed thoroughly by continuously stirring for several minutes. 

Cells were made with Indium-Tin-Oxide (ITO) coated glass plates of size 15 mm $\times$ 10 mm. Polyamides, PI2555 (HD MicroSystems) and JALS 204 (JSR Corporation) were spin-coated on the ITO plates to obtain planar (or homogeneous) and homeotropic alignment of the samples, respectively. These coatings were cured at an appropriate temperature. To make planar cells, baked plates were machine-rubbed unidirectionally and assembled like a parallel-plate capacitor.  
The cell gap was maintained by mixing transparent silica beads ($\approx$ 10  $\mu$m) with UV curable glue. 
Planar cells were filled by the sample in the nematic phase. Textures of samples were viewed using a polarising optical microscope (NIKON POL100).
The temperature of the sample was controlled using an Instec (HSC402XY) hot stage and MK2000 controller. The birefringence ($\Delta n$) measurement was carried out using a phase modulation technique~\cite{Oakberg1997,Oakberg1996}. The setup for the electrooptical measurements consists of a He-Ne laser ($\lambda=632.8$ nm, Thorlabs),  photoelastic modulator (HINDS PEM 100), a pair of crossed Glans-Thompson polarizers,  photo-detector (HINDS),  multimeter (Keithley, 2000 series) and two lock-in amplifiers (Amtek SR7265 and Standford Research SR830). The setup is capable of measuring optical retardation up to 0.1 nm \cite{Oakberg1996}. For dielectric measurements, the ITO electrodes on each substrate were etched into a circular pattern with a diameter of about 7 mm. An  LCR meter (E4980A, Agilent) was used to measure the dielectric constant in the frequency range of 20 Hz to 2 MHz. The splay elastic constant ($K_{11}$) is calculated from the threshold voltage of the Freedericksz transition. All the instruments were computer interfaced and an appropriate LabView programme was developed to control the experiments.

\section{Results and Discussion}
Phase transitions of all the compounds and their mixtures were studied using a polarising optical microscope (POM) and a differential scanning calorimeter (Mettler Toledo DSC 3). Figures \ref{fig:figure1} (b,d) show typical textures of  N\textsubscript{tb} (thread-like focal conics) and N\textsubscript{F} phases (lens-shaped domains), respectively.  Figures \ref{fig:figure2}(a-e) show some representative textures up to 50 wt\% at different temperatures. They exhibit a typical texture of N\textsubscript{tb} phase having elongated thread-like structures running parallel to the rubbing direction. No nematic to nematic transition is observed in mixtures of 60 wt\% and 70 wt\%  (Fig.\ref{fig:figure2}(f) and \ref{fig:figure2}(g)). Instead direct nematic (N) to crystal transition is noted (see later discussions). In the mixtures of 80 wt\% and 90 wt\%, the nematic to N\textsubscript{F} transition temperature is decreased (Figs.\ref{fig:figure2}(h) and \ref{fig:figure2}(i)). 

\begin{figure}[t]
\centering
\includegraphics[scale=0.37]{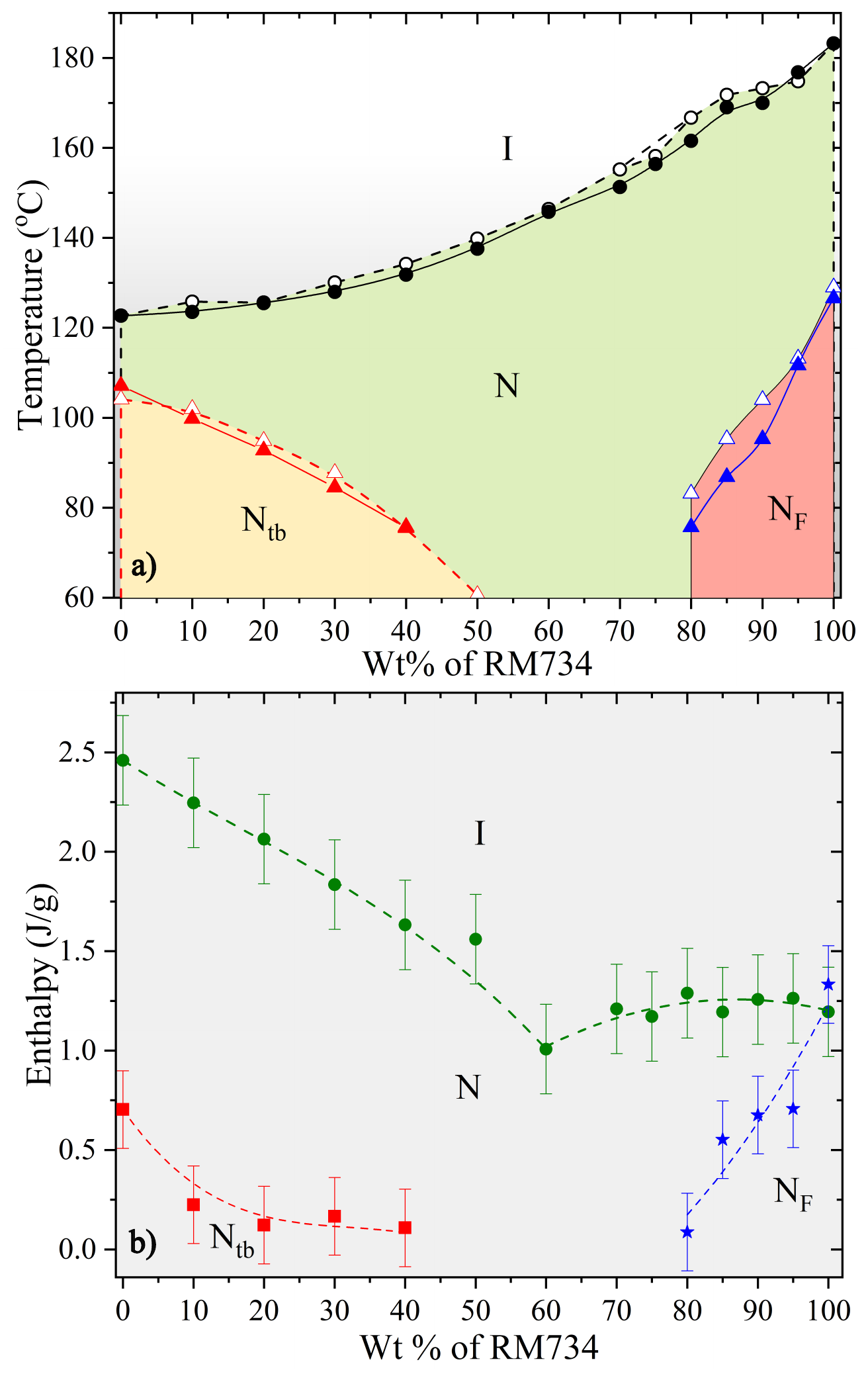}
\caption{(a) Binary phase diagram of CB9CB and RM734. Filled and open symbols represent transition temperatures obtained from DSC and POM, respectively. (b) Enthalpies of N-I transition (green circles),  N-N\textsubscript{tb} transition (red squares), and N-N\textsubscript{F} transition (blue stars). Dashed lines are drawn as a guide to the eye. Error bars represent the standard deviation of the mean.}
\label{fig:figure3}
\end{figure}

  The summary of the observation is presented in Fig. \ref{fig:figure3}(a). The filled and unfilled symbols represent the phase transition temperatures obtained from the differential scanning calorimetry (DSC) and the polarising optical microscopy studies, respectively. DSC measurements were made at the cooling or heating rate of 5\textsuperscript{o}C/min whereas, POM studies were done at 1\textsuperscript{o}C/min. Nevertheless, the phase transition temperatures obtained from two different methods agree quite well.
  
    The phase diagram can be divided into three regions. In the low concentration region of RM734 (0-50 wt\%), the nematic to isotropic (N-I) transition temperature increases while the N-N\textsubscript{tb} transition temperature decreases significantly.  For example, at 50 wt\%, the N-N\textsubscript{tb} phase transition temperature came down to about 60$^\circ$C. It is quite low ($\sim48^\circ$C) compared to the N-N\textsubscript{tb} phase transition temperature in pristine CB9CB. In the high-concentration region (80-100 wt\%) the N-N\textsubscript{F} phase transition temperature also decreases notably. For example, at 80 wt\%, the N-N\textsubscript{F} transition temperature decreases nearly by 50$^\circ$C compared to the pristine RM734. In the intermediate concentration region ($\sim$60-70 wt\%) no nematic to N\textsubscript{tb} or N-N\textsubscript{F} transitions are observed either in the POM observation or from the DSC measurements. The shape and the flexibility of the hydrocarbon link of CB9CB molecules play a crucial role in stabilising N\textsubscript{tb} phase. The addition of rigid rod-like molecules (RM734) might affect these factors and reduces the stability of the N\textsubscript{tb} phase. On the other hand, a smaller concentration of CB9CB greatly reduces the polar order hence the stability of the N\textsubscript{F} phase.\\
 
\begin{figure}[t]
\centering
\includegraphics[scale=0.37]{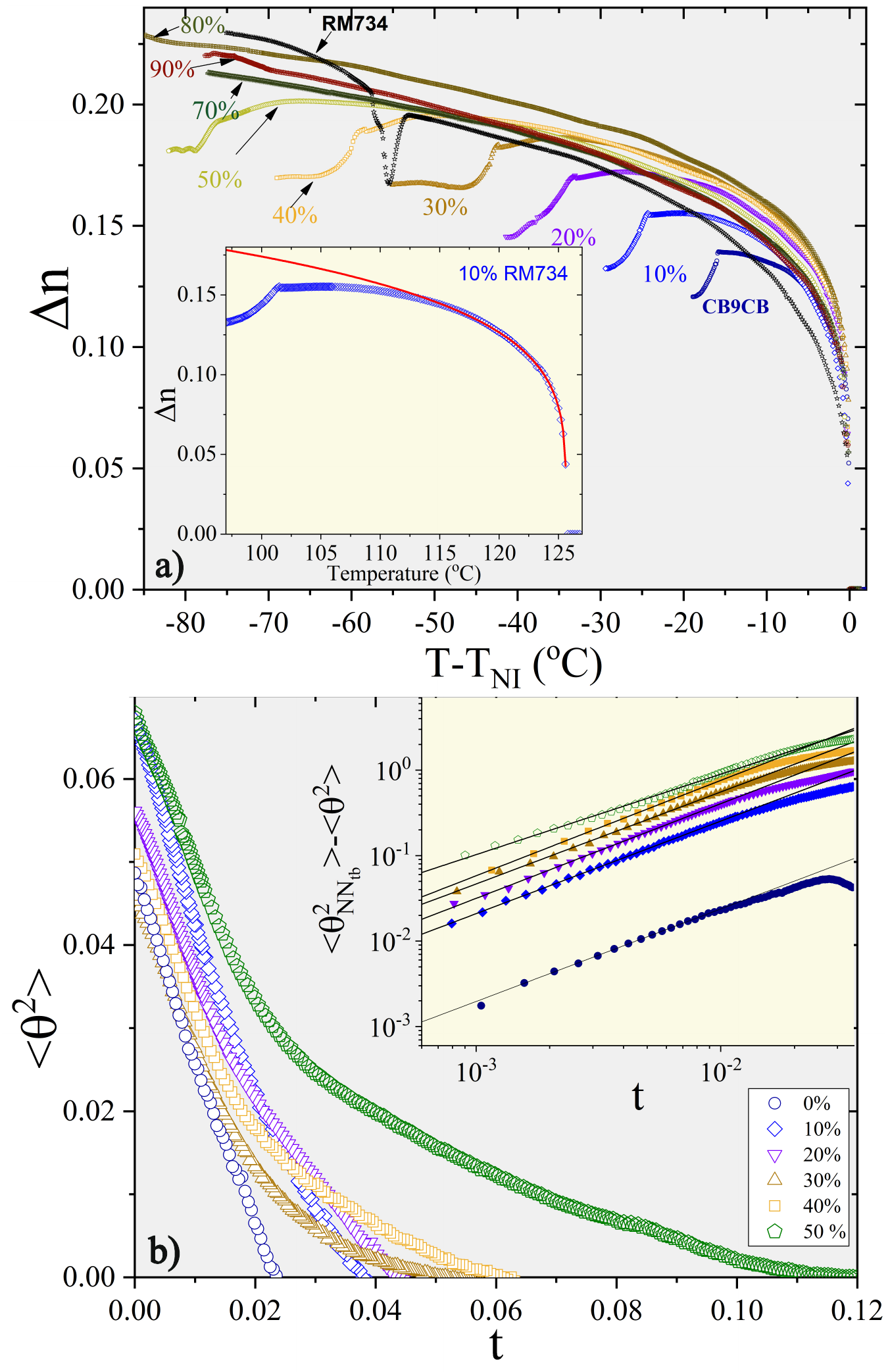}
\caption{(a) Variation of birefringence $\Delta n$ with shifted temperature with increasing wt\% of RM734. Inset shows the variation of $\Delta n$ together with the Haller fit (red line) for 10 wt\% of RM734. (b) Mean-square fluctuations $\langle\theta^2\rangle$ of the tilt
angle calculated from Eq.(2). Inset shows temperature variation of relative tilt fluctuations $\langle\theta_{NN_{tb}}^{2}\rangle- \langle\theta^2\rangle\sim t^{1-\alpha}$ with reduced temperature. For clarity, the critical components of fluctuations are shifted by multiples of 10 along the Y-axis. Cell thickness 9.4 $\mu$m. }
\label{fig:figure4}
\end{figure}

 Figure \ref{fig:figure3}(b) shows the enthalpies of N-I, N-N\textsubscript{tb} and N-N\textsubscript{F} phase transitions of the mixtures including pristine compounds. The enthalpies of the N-I phase transition decrease with increasing wt\% of RM734. The enthalpies of N-N\textsubscript{tb} (0-40 wt\%) or N-N\textsubscript{F} (80-100 wt\%) transitions also decrease substantially with respect to the values of the pristine compounds. In 50 wt\%,  although a clear textural change is observed in the POM, no enthalpy associated with the N-N\textsubscript{tb} transition could be detected in the DSC measurements. It suggests that the N-N\textsubscript{tb} phase transition has become second-order. A similar change of enthalpy was reported in the binary mixtures of CB9CB and 5CB (pentyl cyanobiphenyl)~\cite{CSP_Tripathy_2011}. For 60 wt\% and 70 wt\% the nematic (N) to crystal phase transition occurs below 60$^{\circ}$C (not shown in Fig.\ref{fig:figure3}(b)).
 
The eutectic composition and phase transition temperatures (melting and clearing points)  have been calculated theoretically. We used CSL equations (Le Chatelier-Schr\"oder-van Laar)~\cite{ss2,ss3,ss4,ss5} for predicting the eutectic melting temperature of liquid crystalline mixtures and Cox and Johnson equations for clearing temperature ~\cite{ss2,ss3,ss4,ss5}. The calculated eutectic composition is composed of 83.5 wt\%  CB9CB and 16.5wt\% RM734. The melting and clearing temperatures of the calculated eutectic point are 79.2$^{\circ}$C, and 134.4$^{\circ}$C, respectively. This composition does not match with the composition at which the enthalpy of the NI transition drops suddenly (60 wt\%).  Hence, the enthalpy drop may be associated with poor miscibility of lower-temperature polar phases. 
 
 \begin{table}[b]
\begin{center}
\begin{tabular}{|>{\centering\arraybackslash}m{6cm}|>{\centering\arraybackslash}m{2cm}|}
\hline
Wt\% of RM734 in the mixtures &1-$\alpha ~(\pm 0.02)$\\[2ex]
\hline
0 wt\%  &1.08\\
\hline
10 wt\%  &1.08\\
\hline
20 wt\% &1.1\\
\hline
30 wt\% &1.07\\
\hline
40 wt\% &1.1\\
\hline
50 wt\% &0.95\\
\hline
\end{tabular}
\end{center}
\caption{Critical exponents obtained from different mixtures.}
\label{tbl:Table1}
\end{table}

  \begin{figure*}[!htp]
\centering
\includegraphics[scale=0.38]{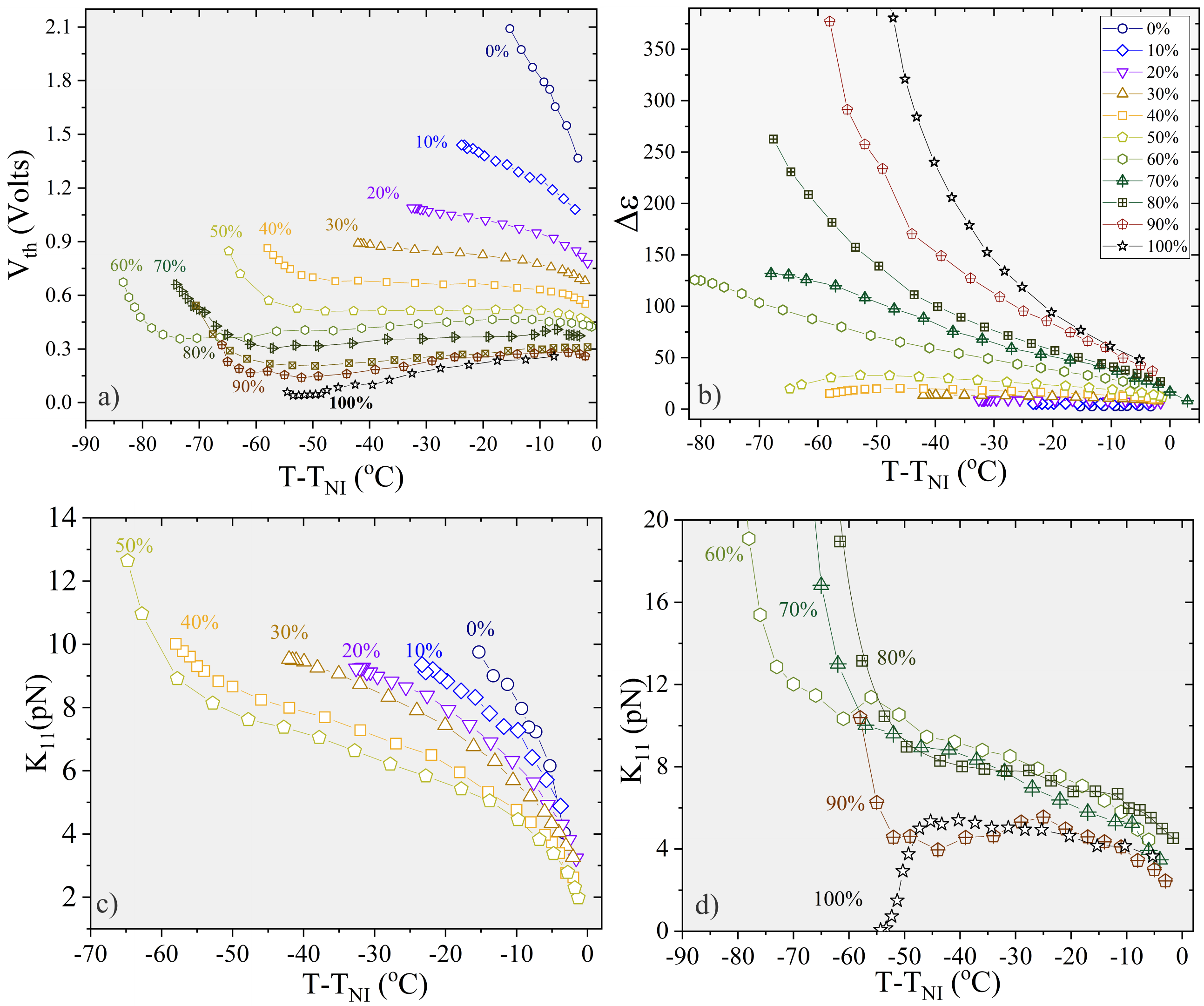}
\caption{ Temperature dependence of (a) Freedericksz threshold voltage $V_{th}$, (b) dielectric anisotropy ($\Delta \epsilon$), (c,d) splay elastic constant $K_{11}$ at different wt\% of RM734. For clarity $K_{11}$ is shown in two different plots. Cell thickness 10.5 $\mu$m.}
\label{fig:figure5}
\end{figure*}  

As a next step, we measured the temperature-dependent birefringence ($\Delta n$) of the mixtures. Figure \ref{fig:figure4}(a) shows the variation of $\Delta n$ of a few mixtures and the pristine compounds. All the samples show a discontinuous jump in $\Delta n$ corresponding to the first-order N-I phase transition followed by a gradual increase with decreasing temperature. An abrupt decrease in  $\Delta n$ at different temperatures for the mixtures (0-50 wt\%) indicates the onset of N-N\textsubscript{tb} phase transition. 
 The temperature-dependent birefringence above the N-N\textsubscript{tb} transition can be fitted to the Haller formula~\cite{Haller1975} 
 \begin{equation}
  \Delta n_p=\Delta n_{max} (1-T/T^{*})^{\beta}
   \end{equation}
   where $\Delta n_{max}$ is the birefringence of the perfectly aligned nematic phase when the orientational order parameter $S=1$ and $\beta$ is the critical exponent connected to the N-I phase transition. A representative fit of Eq.(1) to the birefringence (10 wt\%) is shown in the inset of Fig.\ref{fig:figure4}(a). The exponent obtained for different samples is about $0.2$.
  As the N-N\textsubscript{tb} transition is approached from the high-temperature N phase, the measured birefringence deviates from the prediction of the Haller extrapolation. This deviation is an indication of pre-transitional tilt fluctuations of the director due to the formation of local helical structures in the N-N\textsubscript{tb} transition~\cite{DP}. The fluctuation here is evident up to 50 wt\% of the mixture. The mean-square tilt angle fluctuations $\langle\theta^2\rangle$ can be obtained approximately using the following equation~\cite{Lim_Ho_1978}.
\begin{equation}
  \Delta n_p(T)=\Delta n_{0} (1-\frac{3}{2}\langle\theta^2\rangle)
 \end{equation}	
 where  $\Delta n_p$ is the measured value and $\Delta n_{0}$ is obtained from the extrapolated birefringence (see Eq.(1)) in the absence of fluctuations. Figure \ref{fig:figure4}(b) shows variation of $\langle\theta^2\rangle$ as a function of reduced temperature $t=\frac{T-T_{NN_{tb}}}{T_{NN_{tb}}}$, where $T_{NN_{tb}}$ is the N to N\textsubscript{tb} phase transition temperature. The critical component of the fluctuation depends on temperature and is theoretically predicted to vary as~\cite{Lim_Ho_1978}
 \begin{equation}
  \langle\theta_{NN_{tb}}^{2}\rangle- \langle\theta^2\rangle\sim t^{1-\alpha}, \hspace{0.5cm} \text{for} \hspace{0.5cm} t>0
  \end{equation}
  where $\langle\theta_{NN_{tb}}^{2}\rangle$ and $\alpha$ are maximum fluctuations at the transition and the exponent of the heat capacity, respectively. The inset of Fig. \ref{fig:figure4}(b) shows the variation of $\langle\theta_{NN_{tb}}^{2}\rangle- \langle\theta^2\rangle$ with $t$. The critical range of tilt fluctuations increases with the increasing wt\% of RM734 as the temperature range of the apolar nematic (N) phase increases with concentration. Table~\ref{tbl:Table1} shows that $1-\alpha$ is closer to $1$ for the samples. Similar critical behaviour of the birefringence was reported in CB7CB and other twist-bend liquid crystals exhibiting a broad temperature range of nematic phase~\cite{Damian_P_2018}.\\

In what follows we measure the splay elastic constant of the high-temperature nematic phase from the Freedericksz transition. Figure \ref{fig:figure5}(a) shows the variation of Freedericksz threshold voltage as a function of shifted temperature. $V_{th}$ of pristine CB9CB is much larger than that of pristine RM734. At any fixed temperature $V_{th}$  decreases gradually as the wt\% of RM734 in the mixture increases. Up to 30 wt\%, $V_{th}$ increases with decreasing temperature as $V_{th}\propto S$, where $S$ is the orientational order parameter which increases with decreasing temperature from the N-I transition. The curvature of $V_{th}$ changes above 40 wt\% of RM734. For example, above 40 wt\%, it tends to diverge at lower temperatures. Estimation of splay elastic constant from $V_{th}$ requires dielectric anisotropy $\Delta\epsilon$ $(=\epsilon_{||}-\epsilon_{\perp})$.  Above 50 wt\% of RM734, $\epsilon_{\perp}$ and $\epsilon_{||}$ were measured in planar and homeotropic cells, respectively. This method can not be used for mixtures below 50 wt\% as the homeotropic alignment of the sample is not uniform. Hence, we measured the voltage-dependent dielectric constant in a planar cell and  $\epsilon_{||}$ was obtained from the extrapolation of the high-voltage data~\cite{BKK,Sathya2011}. The dielectric properties of RM734 are quite complex and also depend on the measuring cell thickness~\cite{D1,D2,D3,AE}. We kept the cell thickness almost fixed ($\simeq$10$\mu$m) for all-dielectric measurements. For each sample, we performed dielectric dispersion studies and selected the dielectric constant in an appropriate frequency that is away from any dielectric relaxation. \\\\
Figure \ref{fig:figure5}(b) shows the variation of $\Delta\epsilon$ for all the samples as a function of temperature. In pristine RM734, the anisotropy is very high and also shows a strong temperature dependence. The anisotropy decreases substantially with decreasing concentration of RM734. The splay elastic constant $K_{11}$ was calculated directly using the formula $K_{11} = \frac{V_{th}^2 \epsilon_{o} \Delta \epsilon}{\pi ^2}$~\cite{deg}. Figures  \ref{fig:figure5}(c) and (d) show the variation of $K_{11}$.  All the samples (except pristine RM734) show an increasing trend of $K_{11}$ with decreasing temperature. Up to 30 wt\%, it shows a usual temperature dependence ($K_{11}\propto S^2)$) (Fig.\ref{fig:figure5}(c)).  Above 30 wt\%, $K_{11}$ changes curvature at a particular temperature and diverges as the temperature is reduced. For example, $K_{11}$ in the range of mixtures 50-90wt\%, shows a steep rise below $T-T_{NI}\approx -50^\circ$C. A similar divergence of $K_{11}$ was reported in a few bent-core compounds~\cite{Sathya2011,SD1}. The temperature-dependence of $K_{11}$ for pure RM734 is completely opposite i.e., it decreases sharply as the N-N\textsubscript{F} transition is approached. Such a decrease is proposed to originate from the strong splay fluctuations.  We did not measure the bend elastic constant to avoid degradation of the sample and the alignment due to repeated exposure to a higher voltage for several hours.   
     

 \section{Conclusion}
 We prepared a binary phase diagram of RM734 and CB9CB liquid crystals.  In the mixture,  N-N\textsubscript{tb} and N-N\textsubscript{F} phase transition temperatures decrease substantially compared to the pristine samples and eventually, these two transitions disappear at some intermediate concentrations. The enthalpy of both N-N\textsubscript{tb} and N-N\textsubscript{F} transitions in the mixtures also decreases, suggesting these first-order transitions becomes progressively weak. About 50 wt\%  of RM734, N-N\textsubscript{tb} phase transition becomes second-order. From 60 wt\% to 70 wt\% of RM734, neither N-N\textsubscript{tb}, nor N-N\textsubscript{F} phase transitions are observed. 
 The temperature range of the critical birefringence fluctuations is widened with increasing concentration of RM734 and the critical exponent $1-\alpha$ in the mixtures remains unaffected. The temperature dependence of the splay elastic constant at higher concentrations  (50-90wt\%) of RM734 is significantly different than that of the pristine compounds.
  A small concentration of CB9CB in pristine RM734 remarkably changed the temperature dependence of the splay elastic constant.   Our results suggest that despite their macroscopic resemblance (both being nematic) these two compounds are incompatible due to microscopic disparity. These results invoke further studies on ferroelectric nematic-based mixtures for rich phase diagrams and new physical properties.  \\
       
\noindent\textbf{Acknowledgments}: SD acknowledges financial support from SERB (Ref. No:SPR/2022/000001). AB acknowledges UGC-CSIR for fellowship. This work was supported by the National Science Centre grant 2019/33/B/ST5/02658.  \\


\begin{thebibliography}{99}


\bibitem{VPN} V. P. Panov, M. Nagaraj, J. K. Vij, Yu. P. Panarin, A. Kohlmeier, M. G. Tamba, R. A. Lewis, and G. H. Mehl, Phys. Rev. Lett.
\textbf{105}, 167801 (2010). 
\bibitem{MCE1} M. Cestari, E. Frezza, A. Ferrarini, G. R. Luckhurst,  J. Mater. Chem. \textbf{21}, 12303 (2011).
\bibitem{MCE2}M. Cestari, S. Diez-Berart, D. A. Dunmur, A. Ferrarini, M. R. de la Fuente, D. J. B. Jackson, D. O. Lopez, G. R. Luckhurst, M. A. Perez-Jubindo, R. M. Richardson, J. Salud, B. A. Timimi, and H. Zimmermann, Phys. Rev. E \textbf{84}, 031704 (2011)
\bibitem{Dch} D. Chen, J. H. Porada, J. B. Hooper, A. Klittnick, Y. Shena, M. R. Tuchbanda, E. Korblova, D. Bedrov, D. M. Walba, M. A.Glaser, J. E. Maclennana, and N. A. Clark, Proc. Natl. Acad. Sci. (USA), \textbf{110}, 15931 (2013).
\bibitem{MCO} M. Copic, Proc. Natl. Acad. Sci. (USA) \textbf{110}, 15855 (2013).

\bibitem{VBO} V. Borshch, Y. K. Kim, J. Xiang, M. Gao, A. Jakli, V. P. Panov,
J. K. Vij, C. T. Imrie, M. G. Tamba, G. H. Mehl, and O. D.
Lavrentovich, Nat. Commun. \textbf{4}, 2635 (2013).
\bibitem{LBE} L. Beguin, J. W. Emsley, M. Lelli, A. Lesage, G. R. Luckhurst,
B. A. Timimi, and H. Zimmermann, J. Chem. Phys. B \textbf{116}, 7940
(2012).
\bibitem{GPA} G. Pajak, L. Longa, and A. Chrzanowska, Proc. Natl. Acad. Sci.
U.S.A. \textbf{115}, E10303 (2018).

\bibitem{RJM1}	R. J. Mandle, S. J. Cowling, J. W. Goodby,  Chem. A Eur. J.\textbf{23}, 14554 (2017).

\bibitem{RJM2}	R. J. Mandle, A. Mertelj, blue Phys. Chem. Chem. Phys \textbf{21}, 18769 (2019).

\bibitem{NHI} Nishikawa H, Shiroshita K, Higuchi H, et al. Material with highly polar order. Adv Mater. \textbf{29}(43), 1702354 (2017).
  
\bibitem{MAE}  A. Mertelj, L. Cmok, N. Sebasti\'{a}n, \textit{et al.},  Phys Rev X. \textbf{8}, 041025 (2018).
  
\bibitem{AMA} A. Manabe, M. Bremer and M. Kraska, Liq. Cryst. \textbf{48}(8), 1079 (2021).

\bibitem{NSE} N. Sebasti\'{a}n, L. Cmok, R.J. Mandle, M. R. de la Fuente, I. D. Olenik, M.\u{C}opi\u{c} and A. Mertelj, Phys. Rev. Lett. \textbf{124}, 037801 (2020). 
\bibitem{XCH} X. Chen, E. Korblova, D. Dong, \textit {et al.}, Proc. Natl. Acad. Sci. (USA), \textbf{117}, 14021 (2020).
\bibitem{MBO} M. Born Sitzungsber, Preuss Akad Wiss. \textbf{30}, 614-650
(1916).

\bibitem{IDO} I. Dozov, Europhysics Lett. \textbf{56}, 247 (2001).

\bibitem{Pra1} M. Praveen Kumar,  P. Kula and Surajit Dhara, Phys. Rev. Materials \textbf{4}, 115601 (2020).

\bibitem{Pra2} M. Praveen Kumar,  J, Karcz, P. Kula and Surajit Dhara, Phys. Rev. Materials \textbf{5}, 115605 (2021).

\bibitem{Bart} A. Barthakur,1 J. Karcz , P. Kula  and Surajit Dhara, Phys. Rev. Materials \textbf{7}, 035603 (2023).
\bibitem{MTM1} M. T. M\'{a}th\'{e}, \'{A}. Buka, A. J\'{a}kli and P. Salamon, Phys. Rev. E \textbf{105}, L052701 (2020).
\bibitem{RB} R. Barboza \textit{et. al.}, Proc. Natl. Acad. Sci. (USA), \textbf{119}, e2207858119 (2022).
\bibitem{MTM2} M T. M\'{a}th\'{e}, B. Farkas, L. P\'{e}ter, \'{A}. Buka, A. J\'{a}kli and P. Salamon, Sci. Rep. \textbf{13}, 6981 (2023).

\bibitem{Pra3} M. Praveen Kumar, J. Karcz, P. Kula, S. Karmakar and Surajit Dhara, Phys. Rev. Applied, \textbf{19}, 044082 (2023).


\bibitem{PRAT}R. Pratibha, N. V. Madhusudana and B. K. Sadashiva, \textbf{288}, 5474 (2000).
\bibitem{BKK} B. Kundu, R. Pratibha, N.V.  Madhusudana, \textbf{99}, 247802 (2007).
\bibitem{CLARK}X. Chen, Z. Zhu, M. J. Magrini, E. Korblova, C. S. Park, M. A. Glaser, J. E. Maclennan, D. M. Walba  and N. A. Clark, \textbf{49}(11), 1531 (2022).


\bibitem{Alenka2021}N. Sebastian, R. J. Mandle, A. Petelin, A. Eremin and A. Mertelj, Liq. Cryst., \textbf{48}(14), 2055 (2021).

\bibitem{Oakberg1997} T. C. Oakberg, Proc. SPIE  {\bf 3121}, 19 (1997).

\bibitem{Oakberg1996} T. C. Oakberg, Proc. SPIE  {\bf 2873}, 17 (1996)

\bibitem{CSP_Tripathy_2011}C. S. P. Tripathi, P. Losada-Perez, C. Glorieux, A. Kohlmeier, M. Tamba, G. H. Mehl and J. Leys, Phys. Rev. E \textbf{84}, 041707 (2011).

\bibitem{ss2}H. Le Chatelier, C. R. Acad. Sci., \textbf {100}, 50 (1885).
\bibitem{ss3} I. Z. Schr\"oder, Z. Phys. Chem., \textbf {11}, 449(1893).
\bibitem{ss4}J. J. van Laar,  Z. Phys. Chem., \textbf {63}, 216 (1908).
\bibitem{ss5}R. J. Cox, J. F. Johnson,``Phase equilibria in Liquid crystal mixtures” IBM J. Res. Develop., \textbf{22}, 51 (1978).

\bibitem{Haller1975}I. Heller, Prog. Solid State Chem. {\bf 10}, 103 (1975).


\bibitem{DP}D. Pociecha, C. A. Crawford, D. A. Paterson, J. M. D. Storey, C.T.Imrie, N. Vaupoti\u{c} and E. Gorecka, Phys. Rev. E \textbf{98}, 052706 (2018).

 \bibitem{Lim_Ho_1978} K. C. Lim and J. T. Ho, Phys. Rev. Lett.,  {\bf 40}, 1576 (1978).
 
 \bibitem{Damian_P_2018}D. Pociecha, C. A. Crawford, D. A. Paterson, J. M. D. Storey, C. T. Imrie, N. Vaupotic, and E. Gorecka, Phys. Rev. E, \textbf{98}, 052706 (2018). 



\bibitem{Sathya2011}P. Sathyanarayana, B.K. Sadashiva and Surajit Dhara, Soft Matter, \textbf{7}, 8556 (2011).


\bibitem{D1}  N. Vaupoti\u{c}, D. Pociecha, P. Rybak, J. Matraszek, M. \u{C}epi\u{c}, J. M. Wolska, and E.  Gorecka,  Dielectric response of a ferroelectric nematic liquid crystalline phase in thin cells, 
Liq. Cryst. (https://doi.org/10.1080/02678292.2023.2180099).

 \bibitem{D2}N. A. Clark, X. Chen, J. E. Maclennan, and M. A. Glaser,  Dielectric spectroscopy of ferroelectric nematic liquid crystals: Measuring the capacitance of insulating interfacial layers. https://arxiv.org/abs/2208.09784. (2022). 
 

\bibitem{D3} R. J. Mandle, N. Sebasti\'{a}n, J. Martinez-Perdiguero, and A. Mertelj, Nat. Commun. \textbf{12}, 4962 (2021). 
\bibitem{AE} A. Erkoreka, J. Martinez-Perdiguero, R. J. Mandle, A. Mertelj, N. Sebasti\'{a}n, J. Mol. Liq., \textbf{387}, 122566 (2023). 


\bibitem{deg}de Gennes, P. G. The Physics of Liquid Crystals. Oxford University Press: Oxford, England (1974).

\bibitem{SD1}P. Sathyanarayana, M. C. Varia, A. K. Prajapati, B. Kundu, V. S. S. Sastry and Surajit Dhara, Phys. Rev. E  \textbf{82}, 050701R (2010).




\end {thebibliography}
\end{document}